\documentstyle[preprint,eqsecnum,aps]{revtex}

\begin{document}

\draft

\title{Spontaneous Orbifold Symmetry Breaking and Generation of Mass Hierarchy}

\author{ Israel Quiros\thanks{israel@uclv.etecsa.cu}}
\address{ Departamento de Fisica. Universidad Central de Las Villas. Santa Clara. CP: 54830 Villa Clara. Cuba }

\date{\today}

\maketitle

\begin{abstract}

A very simple mechanism is proposed that stabilizes the orbifold geometry within the context of the Randall-Sundrum proposal for solving the hierarchy problem. The electro-weak TeV scale is generated from the Planck scale by spontaneous breaking of the orbifold symmetry.

\end{abstract}


M-theory represents the most remarkable theoretical success of the end of the millenium. The moduli space of this theory contains all five, anomaly free, ten-dimensional superstring theories and the eleven-dimensional supergravity. In this context the most successfull cosmological construction is that of Horava and Witten\cite{hw}. Compactification of the Horava-Witten theory on an eleven-dimensional orbifold $R^{10}\times S^1/Z_2$ to four dimensions on a deformed Calabi-Yau manifold yields that the resulting theory has $N=1$ supersymmetry. This implies, in turn, that the early universe has undergone a phase where it was five-dimensional\cite{lwc}. In Ref.\cite{lukas} an effective five-dimensional theory was derived by direct compactification of the Horava-Witten theory on a Calabi-Yau space. A static solution to the field equations of this theory exists that may be interpreted as a pair of parallel three-branes that are located at the fixed points of the circle that represent the boundaries of the orbifold $S^1/Z_2$.

A set-up with two three-branes located at the boundaries of a five-dimensional $AdS_5$ spacetime has been used recently by Randall and Sundrum to address the hierarchy problem\cite{randall}. They proposed an scenario where the metric is not factorizable. The four-dimensional metric, in this scenario, is multiplied by a "warp" factor which is a function of the additional dimension

$$
ds^2=e^{-2k|y|}\eta_{nm}dx^n dx^m+dy^2,
\eqno{(1)}
$$
where $k$ is a scale of order the Planck scale, $x^a$ are the usual 4-d coordinates and $0\leq y\leq\pi\lambda$ is the coordinate of the extra dimension. This line-element is consistent with orbifold symmetry ($y\rightarrow -y$) and with four-dimensional Poincare invariance\cite{randall}. Randall and Sundrum have shown that this metric is a solution to Einstein's equations in a simple set-up with two three-branes and appropriate cosmological terms. The two three-branes are located at the orbifold fixed points $y=0$ and $y=\pi\lambda$. These represent the boundaries of the five-dimensional spacetime. 

Working out the consequences of the localized energy density inherent to the brane set-up, Randall and Sundrum found a new solution to the hierarchy problem. In the Randall-Sundrum proposal four-dimensional mass scales are related to five-dimensional input mass parameters and the "warp" factor, $e^{-2k|y|}$. This small exponential factor is the source of the large hierarchy between the observed Planck and weak scales\cite{randall}. In this set-up the relation between the Planck scale and the fundamental scale is found to be

$$
M^2_{Pl}=2M^3\int^{\pi\lambda}_0 dy\;e^{-2k|y|}=\frac{M^3}{k}(1-e^{-2k\pi\lambda}),
\eqno{(2)}
$$
and $M^2_{Pl}$ is a well-defined value even in the limit $\lambda\rightarrow\infty$, in contrast to the product-space expectation that $M^2_{Pl}=M^3 \lambda\pi$. By taking the second "regulator" brane at infinity and considering the coordinate $y=0$ to be the location of the Planck brane, one can derive\cite{lykken}

$$
M^2_{Pl}=2M^3\int^{\infty}_0 dy\;e^{-2k|y|}=\frac{M^3}{k},
\eqno{(3)}
$$
so that if $M$ and $k$ are of order $M_{Pl}=10^{19}GeV$, the graviton zero mode is coupled correctly to generate four-dimensional gravity. For a brane located a distance $y_0$ from the Planck brane we have $M_{Pl}\;e^{-k|y_0|}\sim TeV$, i. e., the electroweak scale (of order TeV) is reproduced by physics confined to the brane located a distance $y_0$ from the brane where the graviton is localized. The generation of this hierarchy requires an exponential of order 30. The advantage of taking a five-dimensional spacetime with infinite extension in the $y$-direction is that one has a better chance of addressing issues such as the cosmological constant problem and black-hole physics\cite{lykken}.

In this letter we further exploit the ideas of Randall and Sundrum. We present a very simple machanism of orbifold geometry stabilization that allows determining the location of the TeV brane. It is literally a mechanism of spontaneous orbifold symmetry breaking. Our starting point is a generalization of the Randall-Sundrum 5-d line-element in the form

$$
ds^2=e^{-2k\sigma(y)}g_{mn}(x)dx^m dx^n+e^{l\gamma(y)+h(x)}dy^2,
\eqno{(4)}
$$
where $l$ is an arbitrary constant factor, $\sigma$ and $\gamma$ are arbitrary functions of the additional $y$-coordinate, while the four-dimensional metric $g_{ab}$ and $h$ are functions of the familiar 4-d coordinates $x^a$. For our purposes it suffices to study a general five-dimensional spacetime without fixing the matter content. Neither standard model nor hidden matter sectors are specified. Therefore, we do not pretend to give any realistic particle picture, but a general five-dimensional gravity set-up. The effective 5-d action we shall study is,

$$
S=\int d^5 x\sqrt{-G}(2M^3 R-\Lambda),
\eqno{(5)}
$$
where $G$ is the determinant of the five-dimensional metric $G_{AB}$ ($A,B=\overline{0,4}$), $R$ is the Ricci scalar obtained from the 5-d Ricci tensor $R_{AB}$ ($R=G^{MN}R_{MN}$) and $\Lambda$ is a cosmological constant that playes the role of a vacuum energy density in the five-dimensional spacetime. The field equations derivable from (5) are

$$
R_{AB}-\frac{1}{2}G_{AB}R=-\frac{\Lambda}{4M^3}G_{AB}.
\eqno{(6)}
$$

These equations can be split into the following set of equations:

$$
R_{ab}-\frac{1}{2}g_{ab}e^{-2k\sigma}R=-\frac{\Lambda}{4M^3}e^{-2k\sigma}g_{ab},
\eqno{(7)}
$$

$$
R_{44}-\frac{1}{2}e^{l\gamma+h}R=-\frac{\Lambda}{4M^3}e^{l\gamma+h},
\eqno{(8)}
$$
and

$$
R_{a4}=-\frac{3}{2}k\;\sigma'\;h_{,a}=0,
\eqno{(9)}
$$
where the prime denotes derivative with respect to the additional $y$-coordinate. Eq.(9) implies two possibilities; either $\sigma'=0$ ($\sigma=const.$) or $h_{,a}=0$ ($h=const.$). In this letter we shall interested in the 2nd possibility and we shall set $h=const.=0$ so the line-element (4) can be written as,

$$
ds^2=e^{-2k\sigma(y)}g_{mn}(x)dx^m dx^n +e^{l\gamma(y)}dy^2.
\eqno{(10)}
$$

Therefore the field equations (7) and (8) yield

$$
^4R_{ab}-\frac{1}{2}g_{ab}\;^4R=3k\;e^{-2k\sigma-l\gamma}(\sigma''-\frac{l}{2}\gamma'\sigma'-2k\;\sigma'^2)g_{ab}-\frac{\Lambda}{4M^3}e^{-2k\sigma}g_{ab},
\eqno{(11)}
$$
and

$$
\sigma'^2=\frac{e^{l\gamma}}{12k^2}(e^{2k\sigma}\;^4R-\frac{\Lambda}{2M^3}),
\eqno{(12)}
$$
respectively.$\;^4R_{ab}$ refers to the familiar four-dimensional Ricci tensor made out of the 4-d Christoffel symbols $\{^{\;a}_{bc}\}=\frac{1}{2}g^{an}(g_{bn,c}+g_{cn,b}-g_{bc,n})$ and $^4R=g^{mn}\;^4R_{mn}$ is the four-dimensional Ricci scalar. The 5-d and 4-d Ricci scalars are related through $R=e^{2k\sigma}\;^4R+2k\;e^{-l\gamma}(4\sigma''-2l\gamma'\sigma'-10k\sigma'^2)$. Combining the trace of Eq.(11) with Eq.(12) one obtains:

$$
\sigma''-\frac{l}{2}\gamma'\sigma'-k\sigma'^2=\frac{\Lambda}{24kM^3}e^{l\gamma}.
\eqno{(13)}
$$

For further simplification of our analysis let us set $g_{ab}=\eta_{ab}$- the usual four-dimensional Minkowski metric. In this case $^4R_{ab}=\;^4R=0$ so, from Eq.(11), one gets

$$
\sigma''-\frac{l}{2}\gamma'\sigma'-2k\sigma'^2=\frac{\Lambda}{12kM^3}e^{l\gamma}.
\eqno{(14)}
$$

Combining Eqs.(13) and (14) yields

$$
\sigma''=\frac{l}{2}\gamma'\sigma',
\eqno{(15)}
$$
which, after integrating, gives

$$
\sigma'=Ce^{\frac{l}{2}\gamma},
\eqno{(16)}
$$
where $C$ is some integration constant. If we put Eq.(15) into Eq.(13) or (14), we obtain

$$
\sigma'^2=-\frac{\Lambda}{24k^2 M^3}e^{l\gamma},
\eqno{(17)}
$$
so the integration constant $C=\sqrt{-\frac{\Lambda}{24k^2 M^3}}$. Eq.(17) (or Eq.(16)) makes sense only for negative $\Lambda<0$ (i. e., $C^2>0$ and the constant $C$ is real). This leads that our model five-diemnsional spacetime is $AdS_5$.

If we introduce a new coordinate through $\frac{d\sigma}{\sigma'}=dy$, it is encouraging noting that the line-element (10) can then be written as

$$
ds^2=e^{-2k|\sigma|}\eta_{mn}dx^m dx^n+r_c^2 d\sigma^2,
\eqno{(18)}
$$
where $r_c=\sqrt{\frac{24k^2 M^3}{-\Lambda}}$ and the orbifold symmetry $\sigma\rightarrow -\sigma$ has been taken into account. While deriving Eq.(18) we have used the following chain of equalities: $e^{l\gamma}dy^2=\frac{e^{l\gamma}}{\sigma'^2}d\sigma^2=\frac{24k^2 M^3}{-\Lambda}d\sigma^2$. The line-element (18) exactly coincides with that of Randall and Sundrum if we set $\Lambda=-24k^2 M^3$.

The most interesting feature of our set-up is contained in Eq.(17) (or Eq.(16)). Since both $\sigma$ and $\gamma$ are functions of the coordinate $y$, Eq.(17) can be written in the general form

$$
\sigma'^2+U(\sigma)={\cal E},
\eqno{(19)}
$$
where ${\cal E}$ is an arbitrary positive constant and $U(\sigma)$ is an arbitrary function of $\sigma$. Then Eq.(19) implies that (see Eq.(17))

$$
e^{l\gamma}=r_c^2[{\cal E}-U(\sigma)].
\eqno{(20)}
$$

Therefore, any solution $\sigma(y)$ of the differential equation (19) in the form

$$
\pm\int\frac{d\sigma}{\sqrt{{\cal E}-U(\sigma)}}=y+C_1,
\eqno{(21)}
$$
implies, in virtue of Eq.(20), a solution $\gamma(y)$. This feature enables us some freedom in the choice of the function $U(\sigma)$. This is, precisely, the feature we shall exploit in searching for a mechanism of orbifold geometry stabilization. In fact, Eq.(19) may be given the following particle interpretation: it represents a scalar $\sigma$-particle with kinetic energy $\sigma'^2$ and total energy ${\cal E}$, that moves along the $y$-direction in a potential $U(\sigma)$. In other words this can be put as follows. The derivative of Eq.(20) yields

$$
l\gamma'=-\frac{dU}{d\sigma}\sigma'/({\cal E}-U),
\eqno{(22)}
$$
so, if we put Eq.(22) into Eq.(15) and, taking into account Eq.(19), one gets

$$
\sigma''=-\frac{1}{2}\frac{dU}{d\sigma}.
\eqno{(23)}
$$

This equation of motion can also be obtained with the help of the variational principle from the action $S_y=\int dy {\cal L}[\sigma',\sigma]$, where ${\cal L}[\sigma',\sigma]=\sigma'^2-U(\sigma)$.\footnotemark\footnotetext{From Eq.(23) one sees that the potential should respect orbifold symmetry, i. e., $U(\sigma)=U(-\sigma)$} Solutions to the equation of motion (23) that correspond to states of least energy are those with $\sigma=\sigma_i$ such that $U(\sigma_i)$ is a minimum. For these solutions the classical hamiltonian $H\sim\int dy[\sigma'^2+U(\sigma)]$ is a minimum too. Therefore, solutions to Eq.(23) for which $\sigma=\sigma_i$ represent the ground $\sigma$-states of the system. These ground states stabilize the orbifold geometry in the sense that the points $\sigma=\sigma_i$ correspond to stable (ground) configurations of the scalar field $\sigma$, yielding that branes located at these points are stable against small perturbations of $\sigma$.

One instructive example is provided by the potential $U(\sigma)=\lambda\sigma^2$, where $\lambda$ is an arbitrary constant. For positive $\lambda>0$ Eq.(21) can be readily integrated to give

$$
\sigma(y)=\sqrt{\frac{{\cal E}}{\lambda}}\sin \sqrt{\lambda}y,
\eqno{(24)}
$$
where $-\frac{\pi}{2\sqrt{\lambda}}\leq y\leq\frac{\pi}{2\sqrt{\lambda}}$, so $-\sqrt{\frac{{\cal E}}{\lambda}}\leq\sigma\leq\sqrt{\frac{{\cal E}}{\lambda}}$ (we have set $C_1=0$). In this case the five-dimensional spacetime has two boundaries that are located at $\sigma=-\sqrt{\frac{\cal E}{\lambda}}$ and $\sigma=\sqrt{\frac{\cal E}{\lambda}}$ respectively. However these are not stable configurations in $\sigma$. The only stable configuration may be localized at $\sigma=0$. For negative $\lambda<0$, we would have, instead,

$$
\sigma(y)=\sqrt{\frac{{\cal E}}{-\lambda}}\sinh\sqrt{-\lambda}y,
\eqno{(25)}
$$
where now $-\infty\leq y\leq\infty$ and $-\infty\leq\sigma\leq\infty$ so, the $\sigma$-direction is unbounded. In this case there is no stable configuration supporting four-dimensional physics. We recall that this is just an instructive example.

A potential that supports the Randall-Sundrum mechanism for generating the mass hierarchy is the following

$$
U(\sigma)=\lambda(\sigma_0^2-\sigma^2)^2,
\eqno{(26)}
$$
where $-\infty\leq\sigma\leq\infty$, i. e., we have an orbifold of $AdS_5$ geometry with infinite extent in the $\sigma$-direction. Following Ref.\cite{lykken} we locate the Planck brane at the origin of the $\sigma$-coordinate $\sigma=0$. The constant $\sigma_0$ is taken in such a way that $e^{-k|\sigma_0|}=TeV/M_{Pl}$ ($k\sim 10^{19}GeV$). For positive $\lambda>0$, the potential $U(\sigma)$ in Eq.(26) has two minimuma at $\sigma=\sigma_0$ and $\sigma=-\sigma_0$ respectively. The origin $\sigma=0$ is a local maximum. In other words, the stable ground $\sigma$-states are located at the minimuma $\sigma=\pm\sigma_0$.\footnotemark\footnotetext{For negative $\lambda<0$, instead, this potential has two maxima at $\sigma=\pm\sigma_0$ and a local minimum at $\sigma=0$. This theory has no bound $\sigma$-states since there may be tunneling from the region $|\sigma|\leq\sigma_0$ into the unbounded region $|\sigma|>\sigma_0$} In this case the electroweak TeV scale is generated from the Planck scale by spontaneous orbifold symmetry breaking. In fact, if the system was initially at $\sigma=0$ (the Planck brane) it "rolls down" until the ground state at $\sigma=\sigma_0$ or at $\sigma=-\sigma_0$ is reached. Once the ground state is reached, say at $\sigma=-\sigma_0$, the orbifold symmetry inherent to the line-element (18) is not a symmetry of this ground state. 

This mechanism also applies in the case of an arbitrary metric $g_{ab}$ if we make the identification

$$
^4\Lambda\equiv (\frac{\Lambda}{4M^3}-3ke^{-l\gamma}[\sigma''-\frac{l}{2}\gamma'\sigma'-2k\sigma'^2])e^{-2k\sigma},
\eqno{(27)}
$$
where $^4\Lambda$ is a four-dimensional cosmological constant. In this case Eq.(11) can be written as

$$
^4R_{ab}-\frac{1}{2}\;^4R=-\;^4\Lambda g_{ab},
\eqno{(28)}
$$
therefore, the requirement that the 4-d cosmological constant should be zero ($^4\Lambda=0$) enables our mechanism to work since, in this case, Eq.(14) and the subsequent set of equations (including Eq.(17)) hold true.

What we have proposed here is a very simple mechanism, literally a mechanism of spontaneous orbifold symmetry breaking, that allows stabilization of the orbifold geometry while solving the hierarchy problem a la Randall-Sundrum. However it is not addressed to give a realistic particle picture. A more realistic approach must contain matter sectors. This will be the subject of forthcoming work.

I acknowledge useful conversations with colleagues Rolando Cardenas and Rolando Bonal and MES of Cuba by financial support of this research.

\end{document}